\documentclass{ws-ijmpc}

\begin{document}

\markboth{F. L. Dubeibe \& L. D. Berm\'udez-Almanza}
{Optimal conditions for the numerical calculation of the largest Lyapunov exponent}

\catchline{}{}{}{}{}

\title{OPTIMAL CONDITIONS FOR THE NUMERICAL CALCULATION OF THE LARGEST LYAPUNOV EXPONENT FOR SYSTEMS OF ORDINARY DIFFERENTIAL EQUATIONS}

\author{F. L. DUBEIBE$^{\dagger}$ \& L. D. BERM\'UDEZ-ALMANZA}

\address{{\it Facultad de Ciencias Humanas y de la Educaci\'on, Escuela de Pedagog\'ia y Bellas Artes, \\
Universidad de los Llanos, Villavicencio, Colombia}\\
{\it $^{\dagger}$fldubeibem@unal.edu.co}}

\maketitle

\begin{history}
\received{xx September 2013}
\revised{xx September 2013}
\end{history}

\begin{abstract}

A general indicator of the presence of chaos in a dynamical system is the largest Lyapunov exponent. This quantity provides a measure of the mean exponential rate of divergence of nearby orbits. In this paper, we show that the so-called two-particle method introduced by Benettin \textit{et al.} could lead to spurious estimations of the largest Lyapunov exponent. As a comparator method, the maximum Lyapunov exponent is computed from the solution of the variational equations of the system. We show that the incorrect estimation of the largest Lyapunov exponent is based on the setting of the renormalization time and the initial distance between trajectories. Unlike previously published works, we here present three criteria that could help to determine correctly these parameters so that the maximum Lyapunov exponent is close to the expected value. The results have been tested with four well known dynamical systems: Ueda, Duffing, R\"{o}ssler and Lorenz.

\end{abstract}
\ccode{PACS Nos.: 05.45.-a, 02.60.Cb, 05.45.Pq}

\section{Introduction}\label{sec:Intro}
Lyapunov exponents tell us whether or not two points in the phase space of a dynamical system, that are initially very close together, stay close in the subsequent motion. In other words, they measure the average rate of divergence or convergence of nearby orbits. The exponential divergence of orbits, in a practical sense, implies the lost of predictability of the system, so any system with at least one positive Lyapunov exponent is defined as chaotic.

\begin{center}
\begin{figure*}[ht]
\begin{center}
\includegraphics[scale=0.53]{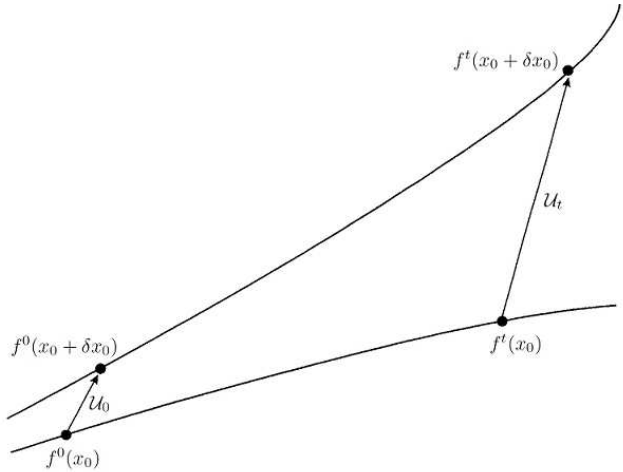}
\end{center}
\caption{Divergence of nearby orbits in phase space.} \label{div}
\end{figure*}
\end{center}

A formal definition can be given by considering the dynamical system
\begin{equation}\label{eq:System}
\dot{x}=F(x),
\end{equation}
where $\dot{x}$ represents the temporal derivative of $x$,\footnote{In all that follows $\dot{q}$ represents the temporal derivative of $q$} with solution $f^{t}(x)$.  Also consider two initial conditions in phase space $x_{0}$ and $x_{0}+\delta x_{0}$, where $\delta x_{0}$ is a small perturbation of $x_{0}$. After time $t$, the solution for the pair of conditions is given by $f^{t}(x_{0})$ and $f^{t}(x_{0} + \delta x_{0})$. Denoting ${\cal{U}}_{t}= f^{t}(x_{0} + \delta x_{0})-f^{t}(x_{0})$, and ${\cal{U}}_{0}= f^{0}(x_{0} + \delta x_{0})-f^{0}(x_{0})=\delta x_{0}$ (see Fig. \ref{div}), the largest Lyapunov exponent $\lambda_{{\rm{max}}}$ is given by\cite{Lyapunov}

\begin{equation}\label{eq:LLE}
\lambda_{{\rm{max}}}= \lim_{t\rightarrow\infty} \frac{1}{t} \ln \frac{|{\cal{U}}_{t}|}{|{\cal{U}}_{0}|},
\end{equation}

As can be noted, there exist as many Lyapunov exponents as phase space dimensions (the so-called Lyapunov characteristic exponents). However, the asymptotic rate of expansion of the largest axis, which corresponds to the most unstable direction of the flow, will obliterate the effect of the other exponents over time making the largest Lyapunov exponent (henceforth LLE) the relevant parameter to determine the degree of chaoticity of the system.

Since the seminal works of Wolf \textit{et al.},\cite{Wolf} Benettin \textit{et al.}\cite{Benettin} and Contopoulos \textit{et al.},\cite{Contopoulos} new methods have been proposed to compute Lyapunov exponents in the literature (see \textit{e.g.} Refs. \refcite{Zeng,Christiansen,Rangarajan,Ramasubramanian,Posch}), some of them for data series and some others for differential equations.

Our main interest in this paper is to propose a solution to the problem of incorrect estimation of LLE when the Benettin \textit{et al.} method (henceforth two-particle method) is used.\cite{Benettin} To do so, we define some criteria that could help to determine optimal parameters of renormalization time and initial distance between trajectories. Additionally, as a reference criterion we will compare the results with the Contopoulos method (henceforth variational method),\cite{Contopoulos} which can be implemented easily and gives very accurate results.

The importance of the two-particle method lies in the fact that in spite of its lack of accuracy, it is a powerful and efficient tool when the system of equations is very cumbersome as in the case of geodesic motion of test particles in General Relativity (see for instance Ref. \refcite{Dubeibe}) or when the linear approximations are not valid, \textit{e.g.} when we are close to a singularity. Moreover, due to the fact that this is an easy to implement method and in some cases the only alternative, in many research fields has been extensively used (see for instance Ref. \refcite{Tancredi} and references therein), a fact that deserves serious attention, taking into account that this method often produces wrong results.

The paper is organized as follow, in section \ref{sec:varmethod} we introduce the variational method which will serve as the reference method. In section \ref{sec:twopmethod} the two-particle method is introduced. Next, in section \ref{sec:false} we present some examples of the incorrect estimation of the LLE for some particular and well known dynamical systems: Ueda, R\"{o}ssler, Duffing and Lorenz. In section \ref{sec:criteria} we propose some criteria for the accurate determination of the largest Lyapunov exponent. Finally in section \ref{sec:conclusion} the conclusions are presented.

\section{Variational Method}\label{sec:varmethod}

Let us consider the dynamical system (\ref{eq:System}), with general solution $f^{t}(x)$ and initial condition $x(t=0)=x_{0}=f^{0}(x_{0})$. The particular solution is given by $x(t)= f^{t}(x_{0})$, so that $\dot{x}(t)= \dot{f}^{t}(x_{0})=F(f^{t}(x_{0}))$. Differentiating the last expression with respect to $x_{0}$,\footnote{In all that follows $D_{q}$ represents the partial derivative with respect to $q$} we get
\begin{equation}\label{eq:var1}
D_{x_{0}}\dot{f}^{t}(x_{0})=D_{x_{0}}F(f^{t}(x_{0}))=D_{x}F(f^{t}(x_{0}))D_{x_{0}}f^{t}(x_{0})
\end{equation}
denoting $D_{x_{0}}f^{t}(x_{0})=\phi^{t}(x_{0})$, equation (\ref{eq:var1}) becomes
\begin{equation}\label{eq:var2}
\dot{\phi}^{t}(x_{0})=D_{x}F(f^{t}(x_{0}))\phi^{t}(x_{0})
\end{equation}
which is the variational equation, with initial condition $\phi^{0}(x_{0})={\cal{I}}$, where ${\cal{I}}$ is the identity matrix.

From the expression for the divergence between nearby orbits, we may write
\begin{equation}\label{eq:ut}
{{\cal{U}}_{t}}= f^{t}(x_{0} + \delta x_{0})-f^{t}(x_{0})= D_{x_{0}}\dot{f}^{t}(x_{0}) \cdot {{\cal{U}}_{0}}
\end{equation}
with
\begin{equation}\label{eq:u0}
{\cal{U}}_{0}= f^{0}(x_{0} + \delta x_{0})-f^{0}(x_{0})=\delta x_{0}
\end{equation}
substituting (\ref{eq:ut}) and (\ref{eq:u0}) into definition (\ref{eq:LLE}), the LLE takes the form\cite{Contopoulos}
\begin{equation}\label{eq:LLEv}
\lambda_{{\rm{max}}}= \lim_{t\rightarrow\infty} \lim_{\delta x_{0}\rightarrow 0} \frac{1}{t} \ln \frac{|\phi^{t}(x_{0})\cdot\delta x_{0}|}{|\delta x_{0}|},
\end{equation}

In order to guarantee that the vector ${{\cal{U}}_{0}}$ have a component in the maximal growth direction, it is very useful to choose an ensemble of $n$ trajectories with different initial orientations, \textit{i.e.},

\begin{equation}\label{eq:LLE-var}
\lambda_{{\rm{max}}}= \lim_{t\rightarrow\infty} \lim_{\delta x_{0}\rightarrow 0} \frac{1}{n}\sum_{i=1}^{n}
\frac{1}{t} \ln \frac{|\phi^{t}(x_{0})\cdot{\delta x_{0 i}}|}{|\delta x_{0 i}|}.
\end{equation}

In essence, in the variational method the LLE is calculated after solving simultaneously the variational equation (\ref{eq:var2}) and the original system (\ref{eq:System}), with initial conditions $x(0)=x_0$ and $\phi^{0}(x_{0})={\cal{I}}$.\cite{Parker}

\section{Two-particle Method}\label{sec:twopmethod}

\begin{center}
\begin{figure*}[ht]
\begin{center}
\includegraphics[scale=0.4]{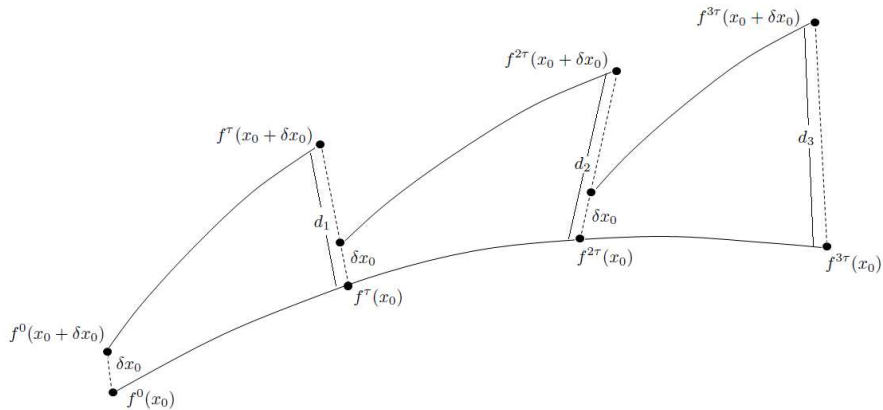}
\end{center}
\caption{Periodic renormalization of the distance for the determination of the LLE in the two-particle method.} \label{2par}
\end{figure*}
\end{center}

 The two-particle method is based on Oseledec's theorem.\cite{Oseledec} In this method we need to consider two trajectories: a reference orbit and a shadow orbit. The reference orbit is solution to the dynamical system (\ref{eq:System}), with initial condition $x(t=0)=x_{0}=f^{0}(x_{0})$, while the shadow orbit is solution to the initial condition $x(t=0)=x_{0}+\delta x_{0}=f^{0}(x_{0}+\delta x_{0})$. After time $\tau$, the distance between the two trajectories is calculated as
\begin{equation}\label{eq:d1}
d_{1}=f^{\tau}(x_{0} + \delta x_{0})-f^{\tau}(x_{0}).
\end{equation}
Then, the point $f^{\tau}(x_{0} + \delta x_{0})$ approaches to the reference orbit along the separation vector $d_{1}$, down to the initial distance $\delta x_{0}$, so that the shadow orbit starts at the same distance for the next iteration (see Fig. \ref{2par}). If this renormalization is made at fixed time intervals $\tau$, then we can write
\begin{equation}\label{eq:di}
d_{i}=f^{i \tau}(x_{0} + \delta x_{0})-f^{i \tau}(x_{0}).
\end{equation}

From definition (\ref{eq:LLE}) an exponent can be calculated for each iteration
\begin{equation}\label{eq:lambdai}
\lambda_{1}= \frac{1}{\tau} \ln \frac{|d_{1}|}{|\delta x_{0}|}, \lambda_{2}= \frac{1}{\tau} \ln \frac{|d_{2}|}{|\delta x_{0}|},  \ldots, \lambda_{i}= \frac{1}{\tau} \ln \frac{|d_{i}|}{|\delta x_{0}|}, \ldots, \lambda_{n}= \frac{1}{\tau} \ln \frac{|d_{n}|}{|\delta x_{0}|},
\end{equation}
such that the LLE is calculated as the average
\begin{equation}\label{eq:LLE-2part}
\lambda_{{\rm{max}}}= \lim_{n\rightarrow\infty} \lim_{\delta x_{0}\rightarrow 0}
\frac{1}{n \tau} \sum_{i=1}^{n} \ln \frac{|d_{i}|}{|\delta x_{0}|}.
\end{equation}

\section{Incorrect estimation of the Lyapunov exponents}\label{sec:false}

\begin{center}
\begin{figure*}[ht]
\begin{center}
\includegraphics[scale=0.5, angle=270]{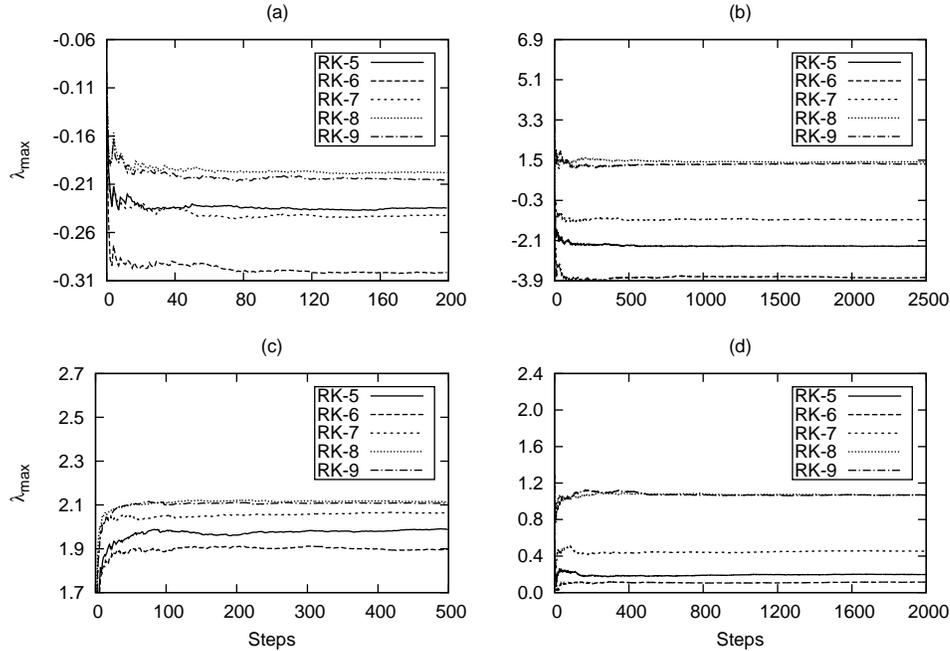}
\end{center}
\caption{LLE calculated with the two-particle method for the systems a) R\"ossler ($\tau=50, \delta x_{0}=10^{-2}$), b) Duffing ($\tau=1, \delta x_{0}=10^{-6}$), c) Lorenz ($\tau=10, \delta x_{0}=10^{-8}$) and d) Ueda($\tau=4, \delta x_{0}=10^{-6}$), using different orders of the explicit Runge-Kutta method. The respective parameters have been set as in table \ref{ta1}.} \label{multi}
\end{figure*}
\end{center}

Both the variational method and the two-particle method, have been widely used in the literature and there is no reason to expect that the results obtained by each method should be different. Yet some authors have pointed out that in many cases the results of both methods are not even similar (see for instance  Ref.~\refcite{Tancredi}). To illustrate this, we shall consider four of the most studied dynamical systems: the Ueda system,\cite{Ueda} R\"{o}ssler system,\cite{Rossler}  Duffing system\cite{Duffing} and Lorenz system.\cite{Lorenz} In table \ref{ta1} we present the systems considered above together with their respective parameters and expected LLE.
\begin{table}[ht]
\tbl{Particular systems with their respective LLE.}
{\begin{tabular}{@{}cccc@{}} \toprule
System & Equations & Parameters & Expected $\lambda$  \\ \toprule
 & $\dot{x}= -y - z \hfill$ &$a$ = 0.15\hfill& \\
R\"{o}ssler \hfill& $\dot{y}= x + a y\hfill$ &$b$ = 0.20\hfill& 0.09 Ref. [\refcite{Wolf}]\\
 & $\dot{z}= b + z (x - c)\hfill $ &$c$ = 10.0\hfill&  \\
\hline
 & $\dot{x}= -\kappa x - y^{3}+A \cos(\omega z)\hfill$ &$\kappa$ = 0.1\hfill& \\
Ueda \hfill& $\dot{y}= x\hfill$ &$\omega$ = 1\hfill& 0.11 Ref. [\refcite{Aguirre}]\\
 & $\dot{z}= 1\hfill $ &$A$ = 11\hfill&  \\
\hline
 & $\dot{x}= \sigma(y - x) \hfill$ &$\sigma$ = 16.0\hfill& \\
Lorenz & $\dot{y}= x (R - z) - y\hfill$ & $R$ = 45.92\hfill& 1.50 Ref. [\refcite{Wolf}]\\
 & $\dot{z}= x y - b z\hfill $ & $b$ = 4.0\hfill&  \\
\hline
 & $\dot{x}= y(1-y^{2})-\alpha x +A\cos(\omega z)\hfill$ &$\alpha$ =0.25 \hfill& \\
Duffing & $\dot{y}= x \hfill$ & $\omega$ = 1.0\hfill& 0.115 Ref. [\refcite{Stefanski}]\\
 & $\dot{z}= 1\hfill $ & $A$ = 0.3\hfill&  \\
\botrule
\end{tabular} \label{ta1}}
\end{table}
In Fig. \ref{multi} we numerically calculate the LLE using the two-particle method with different orders of the explicit Runge-Kutta method\footnote{Henceforth RK-$n$ denotes $n$-th Runge-Kutta order.} by setting arbitrary values of the renormalization time $\tau$, and the initial distance between trajectories $\delta x_{0}$, for each system.\footnote{In all that follows we exclude the 4th Runge-Kutta order. The only reason to do so, is that in most of the cases the numerically calculated value is far apart from the set of values obtained with the higher R-K orders, this behavior force us to increase the range in the vertical axis making the figures unclear for the reader.}

As can be seen from Fig. \ref{multi}, none of these integration methods gives an unique value of LLE in spite of the stable convergence exhibited for long-term evolution. This behavior is not particular for the set of parameters or integration methods used, rather it is a common tendency as pointed out by Tancredi \textit{et al.}\cite{Tancredi} On the other hand, when the variational method is used, the LLE in all cases are practically the same as presented in table \ref{ta1} (R\"ossler 0.088, Ueda 0.108, Lorenz 1.49 and Duffing 0.115).

The possible causes of unreliable estimates of LLE with the two-particle method, have been previously explored by Holman and Murray\cite{Holman} and Tancredi \textit{et al.}\cite{Tancredi} The analysis by Holman and Murray leads to the conclusion that the two-particle technique has an accompanying threshold time scale that depends on the rescaling para\-meters $\delta x_0$ and $\tau$. In other words, after the $i$-th renormalization the distance between trajectories $d_{i}$ is given approximately by
\begin{equation}\label{eq:holman}
d_{i}=\delta x_{0}(1+\alpha \tau^{n})\exp(\lambda_{{\rm{max}}}\tau),
\end{equation}
where $\alpha$ and $n$ are constants associated to the initial power-law transient separation, so that in practice, the numerically calculated LLE is given by
\begin{equation}\label{eq:holman2}
\lambda_{{\rm{max}}} + \frac{\ln(1+\alpha \tau^{n})}{\tau},
\end{equation}
which should affect mainly the quasi-regular trajectories (\textit{i.e.} when $\lambda_{{\rm{max}}} \ll \ln(1+\alpha \tau^{n})/\tau$).

The explanation given by Holman and Murray has been refuted by Tancredi \textit{et al.} who show that $\alpha$ in Eqs. (\ref{eq:holman}) and (\ref{eq:holman2}) is not actually a constant as they assumed, and that the rescaling technique should also lead to a wrong estimate of the LLE when the variational method is used. Furthermore, Tancredi \textit{et al.} found a good agreement in the final values of the LLE for different initial distances up to certain $\delta x_{0}$. With this result, they conclude that there seems to be an optimal value of $\delta x_{0}$ and that the false estimates of the LLE in the two-particle method rely on the accumulation of round-off errors during the computation of the distance between trajectories $d_{i}$ in the course of successive renormalizations.

\begin{center}
\begin{figure*}[ht]
\begin{center}
\includegraphics[scale=1]{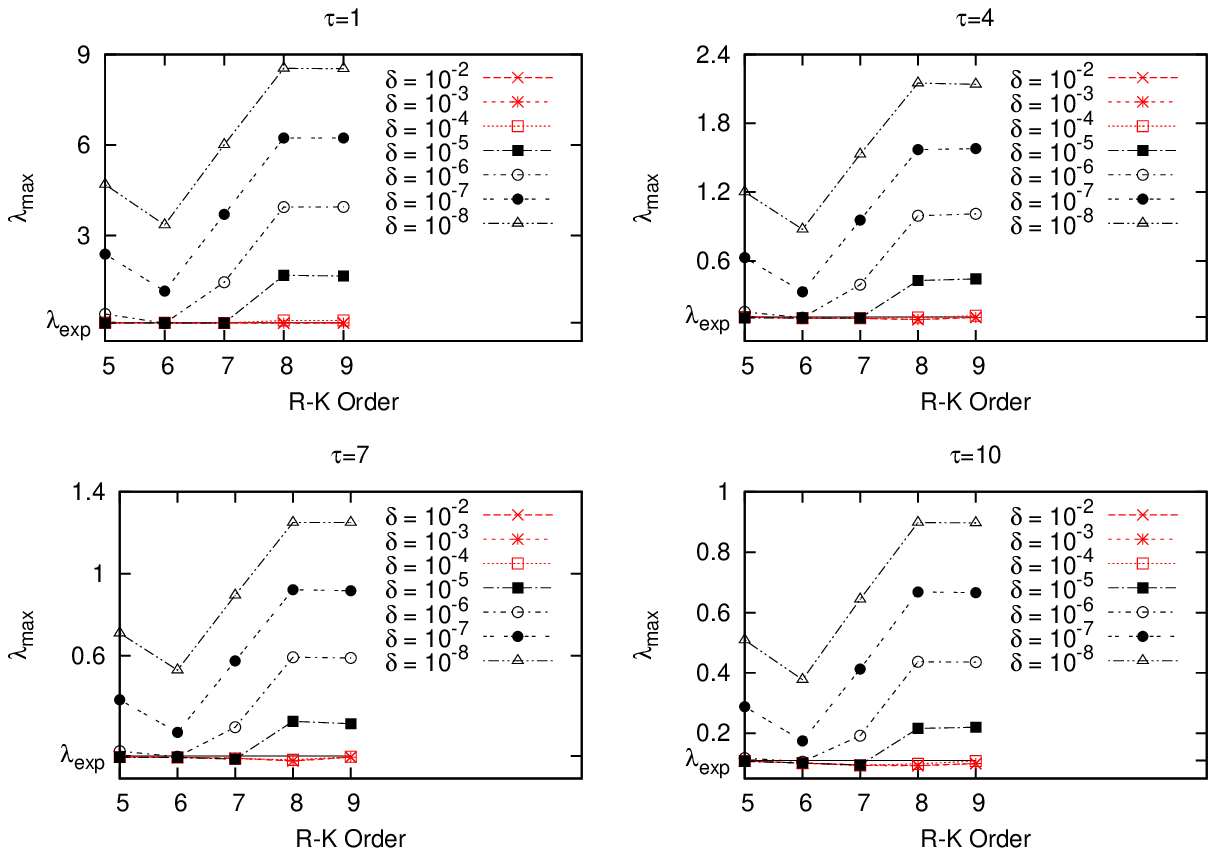}
\end{center}
\caption{(color online) LLE calculated with the two-particle method for the Ueda system using di\-ffe\-rent orders of the explicit Runge-Kutta method, keeping fixed values of the renormalization time $\tau$, for different values of the initial distance between trajectories $\delta x_{0}$. The respective parameters have been set as in table \ref{ta1}. $\lambda_{\rm{exp}}$ represents the expected LLE.} \label{fig:Ueda}
\end{figure*}
\end{center}

\begin{center}
\begin{figure*}[ht]
\begin{center}
\includegraphics[scale=1]{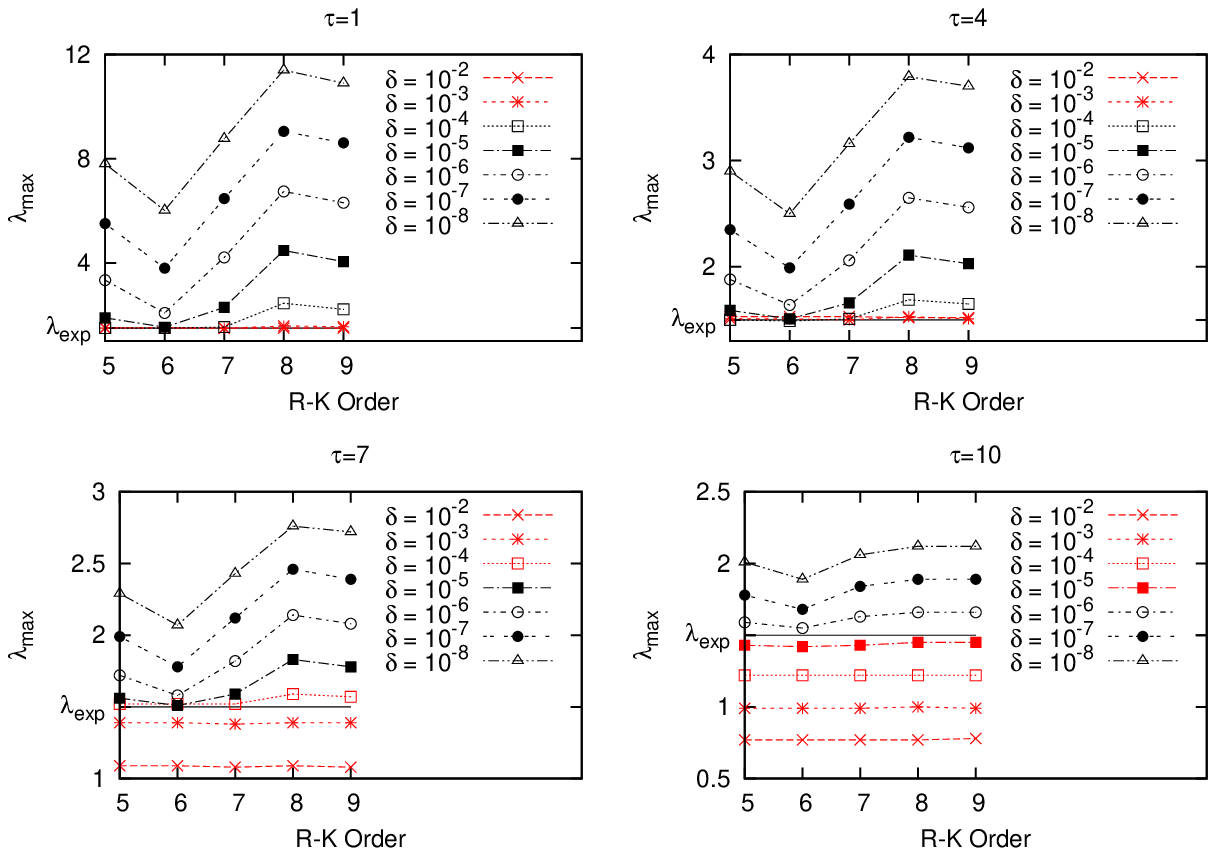}
\end{center}
\caption{(color online) LLE calculated with the two-particle method for the Lorenz system using different orders of the explicit Runge-Kutta method, keeping fixed values of the renormalization time $\tau$, for different values of the initial distance between trajectories $\delta x_{0}$. The respective para\-meters have been set as in table \ref{ta1}. $\lambda_{\rm{exp}}$ represents the expected LLE.} \label{fig:Lorenz}
\end{figure*}
\end{center}

In order to validate (or refute) the premises stated by Tancredi \textit{et al.} and given that the round-off errors should depend on the number of renormalizations, which are indicated by the parameter $\tau$, we start numerically calculating the LLE using the two-particle method with different orders of the explicit Runge-Kutta method for different values of the initial distance between trajectories $\delta x_{0}$, keeping fixed values of the renormalization time $\tau$. The results for the Ueda and Lorenz systems are presented in Figs. \ref{fig:Ueda} and \ref{fig:Lorenz} respectively.

From Fig. \ref{fig:Ueda} it can be seen that when the LLE does not depend on the integration method (red lines), the calculated LLE (which is roughly the expected one) apparently does not depend on the renormalization time $\tau$ nor the initial distance between trajectories $\delta x_{0}$. A different behavior is observed for the Lorenz system Fig. \ref{fig:Lorenz}, in this case even when the LLE does not depend on the integration method (red lines), for a larger renormalization time $\tau$ there is a tendency to a different LLE depending on the initial separation $\delta x_{0}$.

\begin{center}
\begin{figure*}[ht]
\begin{center}
\includegraphics[scale=1]{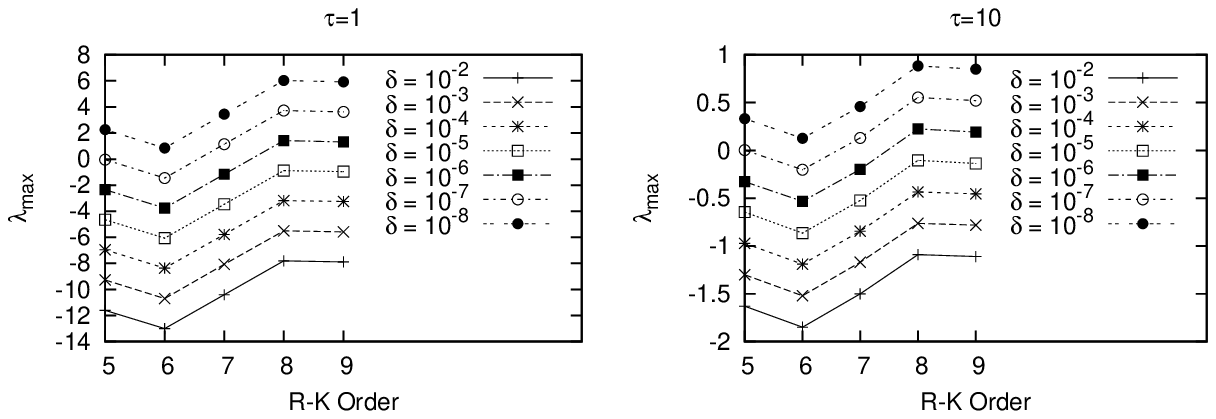}
\end{center}
\caption{LLE calculated with the two-particle method for the Duffing system using different orders of the explicit Runge-Kutta method as indicated in the horizontal axis. The respective parameters have been set as in table \ref{ta1}.} \label{fig:DNW}
\end{figure*}
\end{center}

\begin{center}
\begin{figure*}[ht]
\begin{center}
\includegraphics[scale=1]{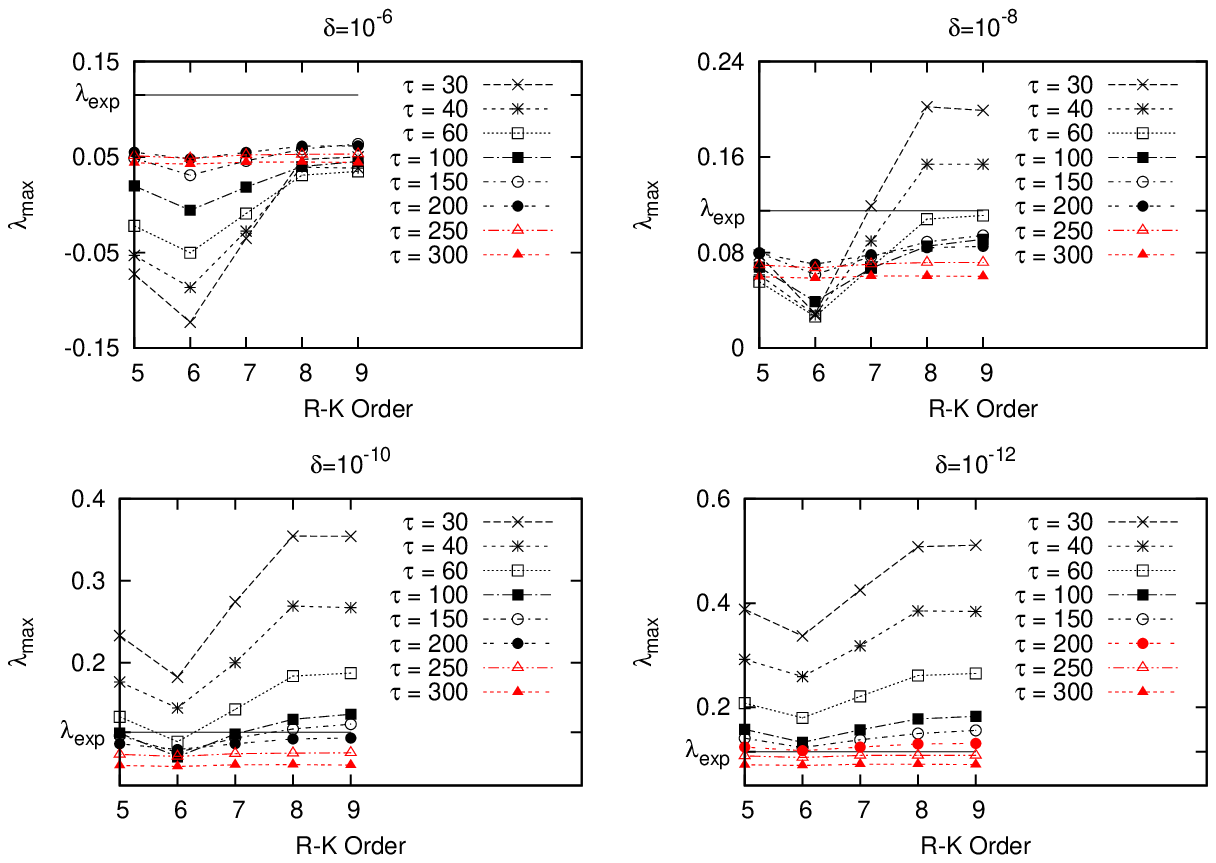}
\end{center}
\caption{(color online) LLE calculated with the two-particle method for the Duffing system u\-sing different orders of the explicit Runge-Kutta method as indicated in the horizontal axis. The respective parameters have been set as in table \ref{ta1}. $\lambda_{\rm{exp}}$ represents the expected LLE.} \label{fig:Duffing}
\end{figure*}
\end{center}

\begin{center}
\begin{figure*}[ht]
\begin{center}
\includegraphics[scale=1]{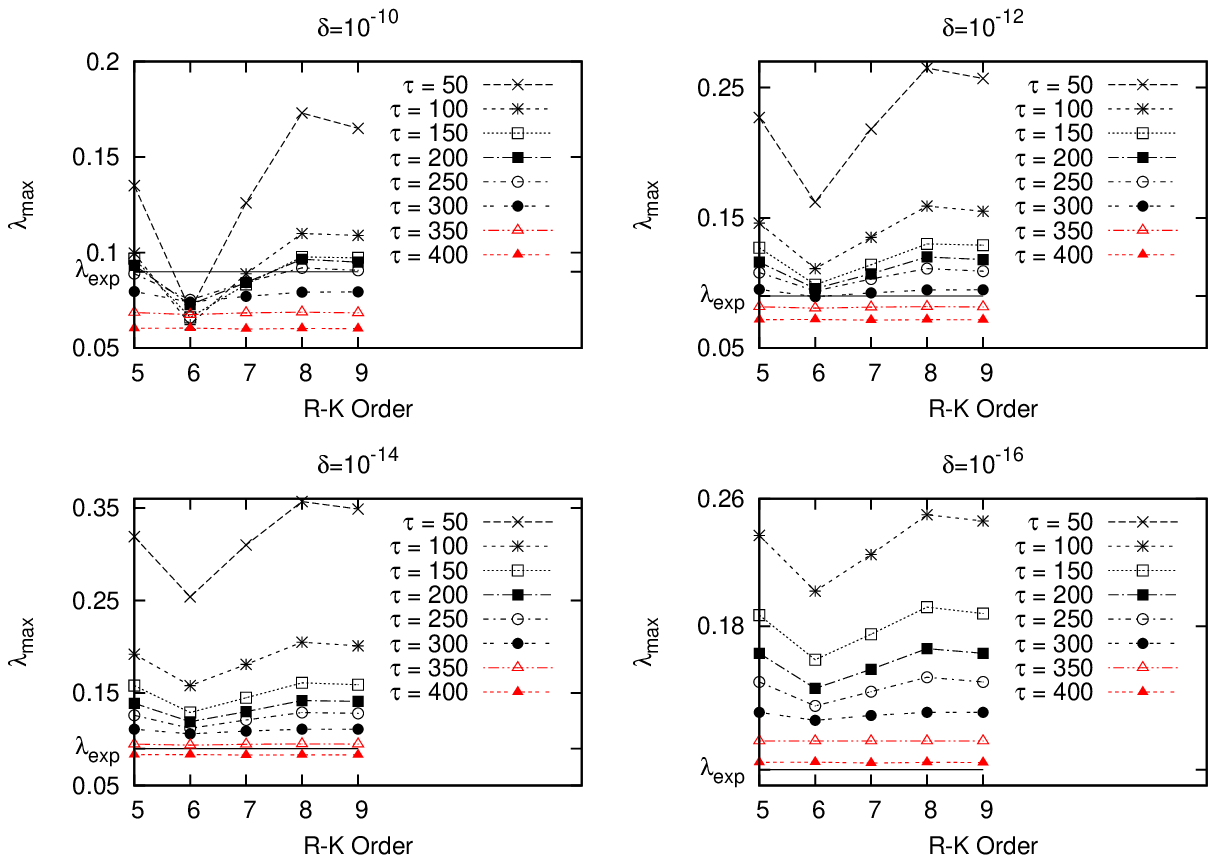}
\end{center}
\caption{(color online) LLE calculated with the two-particle method for the R\"{o}ssler system using different orders of the explicit Runge-Kutta method as indicated in the horizontal axis. The respective parameters have been set as in table \ref{ta1}. $\lambda_{\rm{exp}}$ represents the expected LLE} \label{fig:Rossler}
\end{figure*}
\end{center}

Next we apply the same criteria for the other two systems (Duffing and R\"{o}ssler) for a wide range of $\tau$ values, but in these cases we obtain no tendency towards an unique LLE. The results are presented in Fig. \ref{fig:DNW}. To solve this question we invert the procedure, keeping fixed $\delta x_{0}$ and varying $\tau$. In this case (see Figs. \ref{fig:Duffing} and \ref{fig:Rossler}) we observe a tendency towards unique LLE for certain $\tau$ values (red lines), which is not necessarily the expected LLE. From the plots we conclude that the more accurate result belong to the smaller $\delta x_{0}$,\footnote{Ensuring a $\delta x_{0}$ below the machine precision.} with corresponding smaller $\tau$.

\section{Criteria for the accurate determination of the largest Lyapunov exponent}\label{sec:criteria}

The results presented in the previous section can be explained as follows: The two-particle method could depend on three parameters, namely the number of renormalizations $n$, the initial distance between trajectories $\delta x_{0}$, and the renormalization time $\tau$. When it is possible to guarantee a stable convergence of the LLE, we could ensure that $n$ is large enough to avoid any trouble with the number of terms chosen for the approximation, so the possible incorrect estimates of the LLE should rely on setting $\tau$ and $\delta x_{0}$.

As can be noted from Figs. \ref{fig:Duffing} and \ref{fig:Rossler}, there should exist an optimal range of $\tau$ values. This is due to the fact that if we choose $\tau$ too small it is possible to induce significant round-off errors due to the large number of approximations performed; while choosing too large $\tau$ could cause saturation due to the fact that the chaotic region is generally bounded. Something similar occurs with the $\delta x_{0}$ parameter. From the LLE definition (\ref{eq:LLE}), the distance between trajectories should tend to zero, however from Figs. \ref{fig:Ueda} and \ref{fig:Lorenz} we observe that for small enough $\delta x_{0}$ the calculated LLE depends on the integration method, while choosing a larger value of $\delta x_{0}$ could lead to approximation errors.

The analysis given above and the results of the previous section show us that each particular system can be affected strongly by one or other of the para\-meters, so let us to formulate some simple criteria in order to obtain reliable results of the numerically calculated LLE. This criteria can be stated as follows:
\begin{itemize}
\item The final value of the largest Lyapunov exponent among different runs with different integration techniques should be the same.
\item By fixing $\tau$ and varying $\delta x_{0}$, the largest Lyapunov exponent corresponds to the set of values independent of the integration algorithm, with smaller $\delta x_{0}$.
\item If it is impossible to determine a set of LLE independent of the integration method, we proceed to set a small $\delta x_{0}$ and vary $\tau$, the largest Lyapunov exponent corresponds to the set of identical values with smaller $\delta x_{0}$ and smaller $\tau$.
\end{itemize}

Under these conditions the obtained LLE are close to the expected ones, \textit{i.e.} $\approx 0.12$ for the Ueda system, $\approx 1.44$ for the Lorenz system, $\approx 0.12$ for the Duffing system, $\approx 0.13$ for the R\"{o}ssler system.

\section{Conclusions}\label{sec:conclusion}

In the present paper we have shown that the two-particle method could lead to inconsistent results of the calculation of the largest Lyapunov exponent, particularly when using arbitrary values of the initial separation between trajectories and the renormalization time. With the aim to contribute to the solution of this interesting problem, we performed a numerical exploratory survey which let us propose three criteria that could help to determine confident estimates of the LLE. As shown in section \ref{sec:criteria}, the proposed criteria do not depend of the kind of system under study, and the calculated largest Lyapunov exponent tends to the expected value, independently if it is mainly caused by round-off errors or by approximation inaccuracies. Finally we would like to emphasize that to our knowledge, this is the first proposed procedure to determine optimal values of $\tau$ and $\delta$ in the calculation of the LLE with the method of Benettin {\it et al.}\cite{Benettin}

\section*{Acknowledgments}
We would like to thank the anonymous referee for providing us with constructive comments and suggestions. We enjoyed fruitful discussions with T. Dittrich. During the research for and writing of this manuscript, F. L. D. was partially funded by Universidad de los Llanos under the research project \textit{M\'etodos Num\'ericos para el estudio de la din\'amica de part\'iculas en RG}.

\end{document}